\documentclass[prd, twocolumn, superscriptaddress,floatfix, nofootinbib, preprintnumbers]{revtex4-2}
\usepackage[utf8]{inputenc}
\usepackage[colorlinks=true,citecolor=blue]{hyperref}
\usepackage{amssymb}
\usepackage{amsmath}
\usepackage{graphicx}
\usepackage{footmisc}
\usepackage{url}
\usepackage{multirow}
\usepackage{makecell}
\usepackage{hhline}
\usepackage{comment}
\usepackage{array}
\usepackage{float}
\usepackage{ragged2e}
\usepackage{threeparttable}
\usepackage{enumitem}
\usepackage{xspace}
\usepackage{subfigure}

\usepackage{xcolor}
\usepackage{placeins}

\def\beq{\begin{equation}}
\def\eeq{\end{equation}}
\newcommand{\bea}{\begin{eqnarray}\begin{aligned}}
\newcommand{\eea}{\end{aligned}\end{eqnarray}}

\newcommand{\mycomment}[1]{}

\def\lsim{\mathrel{\rlap{\lower3pt\hbox{\hskip0pt$\sim$}}
   \raise1pt\hbox{$<$}}}         
\def\gsim{\mathrel{\rlap{\lower4pt\hbox{\hskip1pt$\sim$}}
   \raise1pt\hbox{$>$}}}         

\newcommand*{\SoverB}{\ensuremath{S/\sqrt{B}}\xspace}
\newcommand*{\antikt}{anti-$k_t$\xspace}
\newcommand*{\Lambdaprime}{\ensuremath{\Lambda'}\xspace}
\newcommand*{\mq}{\ensuremath{m_{q'}}\xspace}
\newcommand*{\Ekin}
{\ensuremath{E_\mathrm{frac}}\xspace}
\newcommand*{\Ntracks}
{\ensuremath{N_\mathrm{tracks}}\xspace}
\newcommand*{\deltaRsquared}{\ensuremath{\langle(\Delta R)^2\rangle}}
\def\ifb{fb$^{-1}$}
\newcommand*{\pt}{\ensuremath{p_\textrm{T}}\xspace}
\newcommand*{\MeV}{\ensuremath{\text{Me\kern -0.1em V}}}
\newcommand*{\GeV}{\ensuremath{\text{Ge\kern -0.1em V}}}
\newcommand*{\TeV}{\ensuremath{\text{Te\kern -0.1em V}}}

\begin{document}

\title{Soft-unclustered-energy patterns from quirks}

\preprint{DESY-25-081}
\preprint{CERN-TH-2025-119}

\author{David Curtin}
\email{dcurtin@physics.utoronto.ca}
\affiliation{
Department of Physics, University of Toronto, Toronto, Ontario, Canada M5S 1A7
}
\affiliation{Theoretical Physics Department, CERN, 1211 Geneva 23, Switzerland}

\author{Sascha Dreyer}
\email{sascha.dreyer@desy.de}
\affiliation{Deutsches Elektronen-Synchrotron DESY, 22603 Hamburg, Germany}

\author{Max Fust{\'e} Costa}
\email{max.fuste.costa@desy.de}
\affiliation{Deutsches Elektronen-Synchrotron DESY, 22603 Hamburg, Germany}
\affiliation{Institut f\"{u}r Experimentalphysik, Universit\"{a}t Hamburg, 22761 Hamburg, Germany}

\author{Sarah Heim}
\email{sarah.heim@desy.de}
\affiliation{Deutsches Elektronen-Synchrotron DESY, 22603 Hamburg, Germany}
\affiliation{Institut f\"{u}r Experimentalphysik, Universit\"{a}t Hamburg, 22761 Hamburg, Germany}

\author{Gregor Kasieczka}
\email{gregor.kasieczka@uni-hamburg.de}
\affiliation{Institut f\"{u}r Experimentalphysik, Universit\"{a}t Hamburg, 22761 Hamburg, Germany}

\author{Louis Moureaux}
\email{louis.moureaux@cern.ch}
\affiliation{Institut f\"{u}r Experimentalphysik, Universit\"{a}t Hamburg, 22761 Hamburg, Germany}

\author{David Rousso}
\email{david.rousso@desy.de}
\affiliation{Deutsches Elektronen-Synchrotron DESY, 22603 Hamburg, Germany}

\author{David Shih}
\email{shih@physics.rutgers.edu}
\affiliation{NHETC, Department\ of Physics and Astronomy, Rutgers University, Piscataway, New Jersey 08854, USA}

\author{Manuel Sommerhalder}
\affiliation{Institut f\"{u}r Experimentalphysik, Universit\"{a}t Hamburg, 22761 Hamburg, Germany}

\begin{abstract}
  We propose searching for physics beyond the Standard Model in the low-transverse-momentum tracks accompanying hard-scatter events at the LHC. TeV-scale resonances connected to a dark QCD sector could be enhanced by selecting events with anomalies in the track distributions. As a benchmark, a quirk model with microscopic string lengths is developed, including a setup for event simulation. For this model, strategies are presented to enhance the sensitivity compared to inclusive resonance searches: a simple cut-based selection, a supervised search, and a model-agnostic weakly supervised anomaly search with the CATHODE method. Expected discovery potentials and exclusion limits are shown for 140~\ifb\ of 13~\TeV\ proton$-$proton collisions at the LHC.
 
\end{abstract}

\maketitle

\section{Introduction}
In the quest of finding physics beyond the Standard Model (SM) of particle physics at the LHC, it is crucial that no signatures are missed, in particular those involving Soft-Unclustered-Energy Patterns (SUEPs) from low-transverse-momentum (low-\pt) particles. 
Such signatures are challenging, because they often hide under large non-perturbative backgrounds which are hard to model and which make the online trigger selection difficult. 

In the spirit of resonant anomaly detection (for reviews and original references, see~\cite{Kasieczka:2021xcg,Belis:2023mqs,Karagiorgi:2022qnh}), we propose to search for phenomena beyond the SM (BSM) in the low-\pt~tracks ($\lsim$~\GeV) accompanying \TeV-scale resonances. The presence of a resonance allows the online trigger to select high-\pt\ detector objects, and the sidebands of the resonance can be used to estimate backgrounds without having to rely on simulated events. 
 The proposed signature is motivated by a number of theories beyond the SM, some of which predict resonances that are, in fact, de-excited bound states. A typical signature would be a heavy particle$-$antiparticle pair produced in an excited state, which radiates hundreds of low-\pt\ particles as it approaches its ground state, where it finally annihilates, producing a resonant pair of high-\pt\ SM particles.

“Hidden valley” models are an entire class of models beyond the SM with so-far undetected particles that are charged under a new, usually QCD-like force~\cite{Strassler:2006im,Strassler:2008fv}, here referred to as ``dark QCD." The showering and hadronization according to the new force and the subsequent decays can produce a large number of low-\pt~SM particles. Hidden valley models have been discussed in the context of various solutions to the SM problems, such as the dark-matter question~\cite{Hur:2007uz}, and the hierarchy problem. The latter is addressed in theories of neutral naturalness through a twin Higgs scenario~\cite{Chacko:2005pe,Craig:2015pha}, as well as folded supersymmetry~\cite{Burdman:2006tz,Burdman:2008ek,Harnik:2008ax,Curtin:2015fna}, and the Quirky Little Higgs Scenarios~\cite{Cai:2008au}.
These models can give rise to so-called \emph{quirks}~\cite{Kang:2008ea,Harnik:2008ax}, which are dark quarks with a mass that is much larger than the dark QCD confinement scale and would be produced in colliders as quirk$-$antiquirk pairs. In the absence of low-mass dark quarks, the dark color string connecting two pair-produced heavy quirks cannot break but instead supplies a linear restoring force, creating a bound state with the two heavy particles oscillating back and forth on the end of the color string, until they de-excite by various radiative processes and eventually annihilate. 
Here, we study a microscopic quirk scenario where quirk$-$antiquirk pairs that carry SM QCD charge are produced in proton ($pp$) collisions at the LHC and then rapidly de-excite via the isotropic emission of low-\pt\  
SM pions.
Once de-excited, the quirks annihilate into a pair of high-\pt\ jets, as described in Section~\ref{sec:signalmodel}.

We evaluate existing constraints on the developed quirk model from inclusive dijet searches performed by the ATLAS Collaboration, and present strategies to enhance their sensitivity: A simple cut-based selection, a supervised classifier trained on the background estimate and simulated signal events, and a weakly-supervised search for anomalies with the CATHODE framework~\cite{Hallin:2021wme}. Discovery potentials and exclusion ranges are also presented.

Our work demonstrates that the de-exciting pion emission gives rise to a large number of detectable soft tracks that are isotropic in the quirk rest frame, which can serve as a striking discovery signature if it is utilized to extend the reach of a standard dijet search.
We also provide a comparative analysis of various classical and machine learning approaches for the specific class of models under study, i.e. resonances accompanied by track anomalies.

\section{Signal Model}
\label{sec:signalmodel}
A very generic possibility for BSM physics is the existence of a dark QCD sector~\cite{Strassler:2006im}, meaning new states charged under a new asymptotically free gauge force with confinement scale \Lambdaprime. 
If the dark QCD sector has no light dark-quark flavors, which means all dark-QCD-charged matter is much heavier than the dark confinement scale \Lambdaprime, then the lightest dark hadrons are dark glueballs~\cite{Juknevich:2009ji, Juknevich:2009gg} with a mass $m_{GB} \approx 6 \Lambdaprime$~\cite{Athenodorou:2021qvs} for a dark $SU(N)$ confining gauge force. This has important implications for the behavior of any heavy dark-sector states charged under dark QCD.

In any confining gauge theory, if a heavy quark$-$antiquark pair ($q'\overline{q}'$, with $\mq \gg \Lambdaprime$) is produced, it is initially connected by a flux tube of confined force lines that classically would provide a linear restoring force. In SM QCD, this semiclassical behavior does not manifest, since it is energetically more favorable for this flux tube to immediately break by the creation of light quark$-$antiquark pairs with mass below the confinement scale $\Lambda$, corresponding to the production of light mesons in hadronization. 
However, in a dark QCD sector with no light quark flavors, the flux tube cannot break. The semiclassical linear restoring force of a confining gauge theory in the long distance ($r > 1/\Lambdaprime$) regime is therefore manifest, and the dark quark$-$antiquark behaves as though it is connected by a spring, oscillating back and forth with maximal linear separation $l \sim \mq/{\Lambdaprime}^2$ (assuming that the dark quarks are created with kinetic energy $\sim \mq$ in their rest frame). Such dark quarks are called \emph{quirks}.

Upon creation of the ``quirkonium'' state, it will start to oscillate.
The quirks can in principle annihilate each time they pass close to each other, but like any accelerating charges, they also immediately start de-exciting by the emission of radiation. 
This imparts orbital angular momentum onto the quirk pair, which greatly suppresses their annihilation rate, until the quirkonium reaches its ground state. In most scenarios, the quirkonium lifetime is therefore determined by the \emph{de-excitation time scale}.

The type of radiation emitted depends on the quirk quantum numbers. If it is charged under electromagnetism (or any unbroken gauge force, like a dark $U(1)_D$), it will emit (dark) photons via bremsstrahlung. 
If it carries SM QCD charge, the quirks will emit SM pions.
The one guaranteed de-excitation channel is emission of dark glueballs, since the quirk has to be charged under dark QCD.
Quirk de-excitation dynamics is famously complicated. Photon emission is calculable, while the emission of pions can be crudely modeled to be isotropic and approximately thermal with a Hagedorn momentum spectrum~\cite{Kang:2008ea}, provided the string length is much smaller than $\Lambda^{-1}$. 
On the other hand, no rigorous treatment of quirk de-excitation via glueball emission exists. A similar ansatz to pion emission could be applied, but only as long as the quirk kinetic energy remains much larger than the glueball mass. Even so, the absolute glueball emission rate is very difficult to predict.

\begin{figure}
\centering
  \subfigure{\includegraphics[width=0.49\textwidth]{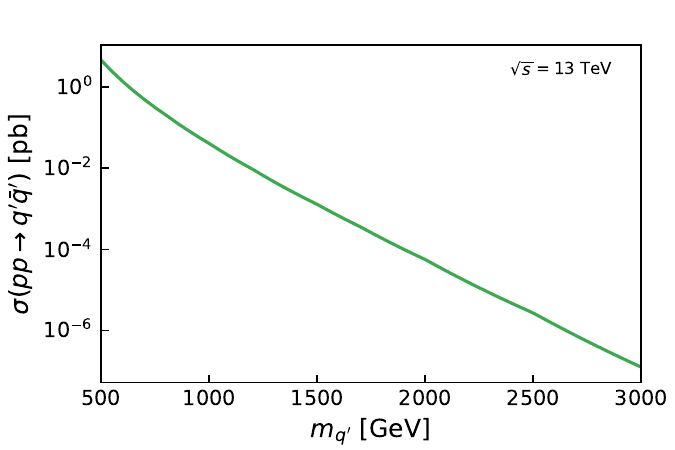}}   
  \subfigure{\includegraphics[width=0.49\textwidth]{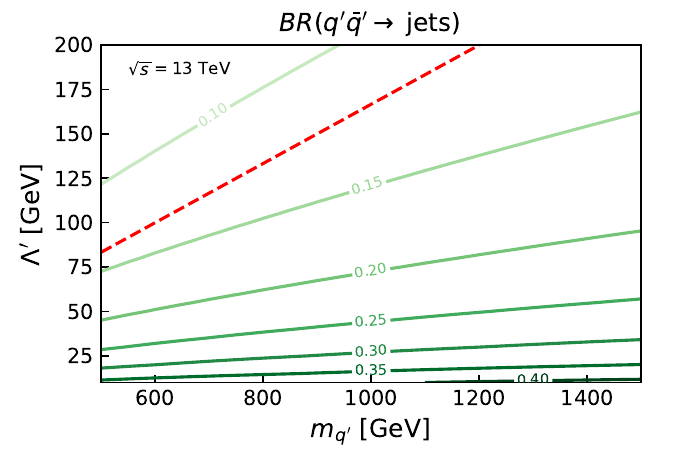}}
   \caption{Cross section of quirk$-$antiquirk production in $pp$ collisions at 13~\TeV\ as a function of the quirk mass (top). Branching ratio for the diquirk system to decay into two SM jets, depending on the quirk mass and \Lambdaprime (bottom).  The dashed red line indicates where quirk annihilation to two dark glueballs is kinematically forbidden, invalidating the presented simplified estimate. Close to the dashed red line, the quirks dominantly or entirely annihilate to SM jets. \label{fig:xs_br}}
\end{figure}

The above illustrates that quirks can exhibit highly exotic behavior~\cite{Burdman:2008ek, Cheung:2008ke, Harnik:2008ax, Kribs:2009fy, Cai:2008au, Craig:2015pha, Nussinov:2014xua, Harnik:2011mv, Fok:2011yc, Knapen:2017kly, Farina:2017cts, Knapen:2016hky, Chacko:2015fbc, Cheng:2015buv, Evans:2018jmd, Li:2020aoq, Li:2019wce, Li:2021tsy, Borsato:2021aum, Forsyth:2025wks, Sha:2024hzq, Feng:2024zgp, Li:2023jrt}. We will focus on the regime where the quirk string is microscopic compared to the SM hadronic scale.
Our signal model consists of quirks charged under both QCD and dark QCD, with masses in the phenomenologically interesting $m_{q'} \sim \mathcal{O}(\mathrm{TeV})$ range for LHC resonance searches, while $\Lambdaprime \gsim \mathcal{O}(10~\mathrm{GeV})$ to ensure the quirk string satisfies $l \sim E_{kin}/{\Lambdaprime}^2 \sim \mq/{\Lambdaprime}^2 < \Lambda^{-1}$~\cite{Kang:2008ea}. Without going into a detailed calculation of pion emission, this also ensures prompt (within a ps) de-excitation and annihilation, even for an implausibly small pion emission probability per oscillation.
We are agnostic as to whether these dark quarks carry electromagnetic charge, since their phenomenology will be determined by their strong interactions. 
For concreteness, we take the dark QCD to be $SU(3)_D$, and the quirks to be fundamentals under $SU(3)_c \otimes SU(3)_D$. 
For their LHC production cross sections (see Fig.~\ref{fig:xs_br}) we can use the next-to-leading order (NLO) results for vector-like quark (VLQ) production in Ref.~\cite{Fuks:2016ftf}, multiplied by a dark color factor of 3. The dark color contribution to the $K$-factor will be a subdominant effect.
After de-excitation, the $q'\overline{q}'$ state can annihilate into QCD and dark-QCD gluons, with a branching ratio to SM jets given by $\mathrm{Br}(q'\overline{q}'\to jj) = \alpha_S^2/(\alpha_S^2+\alpha_D^2)$.\footnote{This assumes that the quirk annihilates in the SM color-singlet state, which is justified for microscopic quirk strings with $l < \Lambda^{-1}$. In analogy to SM heavy quarkonium, annihilation of the de-excited quirk pair will be dominated by the $s$-wave color-singlet state, with the color-octet annihilation being suppressed by $v^2$, where $v$ is the nonrelativistic speed of the quirks in the de-excited bound-state rest frame~\cite{Bodwin:1994jh}.} Both couplings are evaluated at the scale $m_{q'}$, and $\alpha_D(m_{q'})$ is determined, assuming that $q'$ is the lightest dark-QCD charged state in the theory, by one-loop pure Yang-Mills RG-evolution from $\alpha_D^{-1}(\Lambdaprime)=0$. The resulting annihilation branching fraction to QCD jets is shown in Fig.~\ref{fig:xs_br} (bottom). 

The  estimate of the branching ratio to SM jets given above does not take into account the phase space suppression from the dark glueball mass, which can be significant at large $\Lambdaprime$. We therefore also show in Fig.~\ref{fig:xs_br} (bottom) a dashed red line indicating when the lightest dark glueball mass $m_{GB} \approx 6 \Lambdaprime$ exceeds \mq. Above this line, annihilation to two dark glueballs is kinematically forbidden, and the branching ratio to SM jets will be close to 1 (though a subdominant mixed decay to one dark glueball and a SM jet may occur). Computing the precise decay width including glueball mass effects is beyond the scope of this paper, but around this threshold, the SM jet branching ratio of quirk annihilation will be close to 1.

Following $q' \overline{q}'$ pair creation, we model de-excitation of the quirkonium state by converting the quirk kinetic energy $E_{kin}$ in the quirkonium rest frame into isotropic pion emission with a thermal energy spectrum $dN/d^3\mathbf{p} \propto \exp(-E/T_H)$, where $T_H = 1.2 \Lambda$ is the Hagedorn temperature taken to be 150~\MeV.  
This results in the emissions of $N_\pi \sim E_{kin}/T_H \sim \mq/T_H \sim \mathcal{O}(10^3)$ pions per quirk pair, with relative probabilities of pion charges given by isospin ($1/3$ each for positive, negative and neutral). The last emitted pion charge is  dictated by overall charge conservation. See also Ref.~\cite{Harnik:2008ax} for an early discussion of this soft-track quirk signal.

The fraction of quirkonium internal energy that is radiated away as dark glueballs is unknown. In general, the quirks may also have other dark sector interactions that open additional de-excitation channels, such as dark photon emission.
We account for this uncertainty by introducing a parameter 
0.2 $\leq$ \Ekin $\leq$ 1 into our model, setting the fraction of quirkonium internal energy that is converted into thermal isotropic pions.

The final model therefore has three parameters: the mass of the quirks \mq, the energy scale \Lambdaprime, and the fraction of energy going to pions \Ekin. 
In practice, for the range of dark confinement scales we study, the exact value of \Lambdaprime only changes the cross section normalization of the dijet annihilation signal by determining $\mathrm{Br}(q' \overline{q}' \to jj)$.
In order to test the sensitivity, we also introduce the signal strength $\mu$, which scales the cross section of the model with the chosen parameters. For example, a sensitivity to $\mu = 10$ means that we can find the signature, but only if its cross section is 10 times as large as predicted by the nominal model.

\section{Signal and background simulation}
\label{sec:simulation}
Assuming $pp$ collisions at 13~\TeV\ center-of-mass energy, the \textsc{MG5\_aMCatNLO} 3.4.2 ~\cite{Alwall:2014hca} MC generator was used to generate the signal and background events, together with \textsc{Pythia} 8.2~\cite{Sjostrand:2014zea} for parton shower, hadronization and the underlying event from multiple parton interaction (MPI). The NNPDF23 set~\cite{Ball:2012cx} was used. 
No pile-up events were added, as the simplifying assumption is made that the track$-$vertex association reduces the contribution from pileup tracks significantly (to 10\% or less, see Ref.~\cite{ATLAS-CONF-2012-042}) and the remaining contribution can be estimated \cite{ATLAS:2022ctr}. Obviously, the analysis strategy will have to be revisited for the High-Luminosity LHC with its more challenging pile-up conditions but improved tracking detectors.

\subsection{Signal simulation}

The generation of simulated signal samples was performed in a semi-analytical hybrid approach. The simulation consists of three components: the simulation of the quirk$-$antiquirk production, the radiation of pions during the oscillation/de-excitation process, and the decay of the quirkonium. Quirk$-$antiquirk production is approximated using top-quark-like heavy-vector-like quark triplets~\cite{Buchkremer:2013bha,VLQUFO} generated with the \textsc{MG5\_aMCatNLO} MC generator and \textsc{Pythia}. The non-perturbative emissions of the pions during the de-excitation are simplified by treating them as isotropic radiation, thereby bypassing detailed simulation of each radiated particle by assuming an isotropic thermal distribution. Depending on the relative kinetic energy of the two quirks, and the chosen parameter value of \Ekin, charged and neutral pion four-vectors are produced isotropically with an energy sampled from a relativistic Maxwell-Boltzman distribution with a temperature that corresponds to $T_H$.
This thermal approach is inspired by earlier SUEP studies~\cite{Barron:2021btf}. The  momentum of the pion system is adjusted to that of the diquirk system. 
Finally, the quirk$-$antiquirk annihilation to two hadronic jets is produced separately with \textsc{MG5\_aMCatNLO}  from the decay of a heavy scalar particle, whose mass corresponds to twice the quirk mass, also with a total  momentum adjusted to that of the diquirk system. Events were simulated for masses ranging between 750 and 1500~\GeV, and \Ekin values between 0.2 and 1.0.

Note that in general we neglect the momentum imparted on the quirk bound state during de-excitation, and treat total pion emission as exactly isotropic. 
This is a good approximation when de-excitation is dominated by pion emission, i.e. $\Ekin$ is close to 1. Given most quirk de-excitations emit thousands of pions, with an average pion momentum of $\langle p_\pi \rangle \sim T_H$, the total momentum of the emitted pion radiation is roughly  $|\vec p | \sim N_{\pi}^{1/2} \langle p_{\pi} \rangle  \sim (E_{kin} T_H)^{1/2} \sim \mathcal{O}(10~\textrm{GeV})$, which is a very small correction to our event kinematics given the quirk mass scales we study. 
Similar arguments apply when $\Ekin < 1$ and the additional dark radiation is dominated by light states like dark photons. 
The case where $\Ekin < 1$ and the dark radiation is dominated by dark glueballs deserves special mention. 
The dynamics of glueball emission during quirk de-excitation are unknown, but we can write the average number of emitted glueballs to be $N_{GB} \sim \alpha E_{kin}/m_{GB} \sim \alpha E_{kin}/(6 \Lambdaprime)$, with $\alpha < 1$, while each glueball's average momentum is likely $\langle p_{GB} \rangle \sim \beta m_{GB} = 6 \beta \Lambdaprime$, with $\beta < 1$ if emission is Hagedorn-like with an energy set by $T_{H}^\prime \sim \Lambdaprime$.  For $\Ekin \ll 1$, this yields an estimate for the total momentum of the emitted glueball radiation 
of $|\vec p| \sim (N_{GB})^{1/2} \langle p_{GB} \rangle \sim \beta \alpha^{1/2} (E_{kin} \Lambdaprime)^{1/2}$. Given that we study $\Lambdaprime$ up to $\mathcal{O}(100~\mathrm{GeV})$, the resulting momentum shift on the quirk system may not be entirely negligible, and may appreciably shift the kinematics of the resulting dijet signal. However, we neglect this effect for two reasons. First, it is still expected to be subdominant and not greatly impact the reach of the dijet search once the additional pion track criteria are imposed. Second, given the unknown details of de-excitation via glueball emission, we choose to keep the analysis agnostic of those details (thereby letting the results apply to other sources of dark radiation). It would certainly be interesting to repeat our studies for $\Ekin< 1$ at some future time, when the dynamics of de-excitation via glueball emission are better understood.

\subsection{Background simulation}

The dijet background, including MPI, was simulated at leading order with the \textsc{MG5\_aMCatNLO} generator and \textsc{Pythia} 8. A trijet sample was also generated at leading order to confirm that the contribution of events with a third high-\pt jet is negligible even in the tails of the studied distributions. Events were simulated for each of the following dijet invariant mass regions: 1.1$-$1.5~\TeV\ (400M), 1.5$-$2.2~\TeV\ (80M), 2.2$-$4~\TeV\ (10M), 4$-$6~\TeV\ (10M), 6$-$9~\TeV\ (1M). 

\subsection{Preselection and distributions}
\label{subsec:presel_dis}

The generated events were processed through a simplified LHC detector simulation~\cite{RnDCard} based on \textsc{Delphes} 3.5~\cite{deFavereau:2013fsa}.

The applied selection criteria approximate the ATLAS dijet search~\cite{ATLAS:2019fgd}, except that the radius parameter of the selected jets is increased from \antikt R = 0.4 to 1.0, in order to improve the dijet invariant mass resolution through better including final state radiation (FSR). Of course, this also increases the number of pion tracks that are wrongly included in the resonance. The resulting signal width is about 20\% of the resonance mass if approximated by a Gaussian distribution. Without the radius increase, the signal, even with zero additional pion radiation, would be significantly wider~\cite{ATLAS:2019fgd}, causing a reduction in significance and challenges for the CATHODE algorithm.  A dedicated FSR recovery algorithm would, of course, be even more powerful, and more robust against initial state radiation (ISR) and the inclusion of pion radiation. 

Events are required to have at least two jets in the detector acceptance with transverse momentum larger than 150~\GeV. Tracks must be in the detector acceptance and have transverse momenta larger than 0.5~\GeV, the current threshold for track reconstruction in ATLAS~\cite{ATLAS:2023iat}. A selection is applied on half of the rapidity separation between the two leading jets, $y^* = (y_1 - y_2)/2$, where $y_1 (y_2)$ is the rapidity of the (sub-)leading jet. To favor s-channel processes and suppress QCD, only events with $|y^*| < 0.6$ are selected. 
The distribution of the dijet invariant mass $m_{jj}$ is only considered above 1.3~\TeV, as for lower values the required number of simulated background events would be too high. 
The $m_{jj}$ signal region (SR) depends on the probed \mq. Its definition aims to contain as much signal as possible while keeping the window narrow enough for a good signal over background ratio and for the CATHODE algorithm to perform well. Further considerations are the exponentially falling background and the need to accommodate different signal shapes for different values of \Ekin. 
The SR is defined such that 68\% of the signal events for \Ekin = 1.0 are required to fall below its upper bound in $m_{jj}$.  
The lower bound is defined such that the difference between it and 2\mq is half the difference between 2\mq and the upper bound. 
For example, for \mq = 1.5~\TeV, the window is 2.82$-$3.36~\TeV\ to ensure good coverage of the invariant mass peak at at approximately 2\mq = 3~\TeV.

Together with the invariant mass requirement, the $y^*$ selection translates into an implicit selection cut on the transverse momentum of the jets, which is significantly higher than the initial 150~\GeV, such that they would pass the corresponding trigger requirements applied by the experiments. The expected numbers of signal and background events for different mass points and values of \Ekin, for 140~\ifb\ of $pp$ data, corresponding to the LHC Run-2 dataset collected by ATLAS or CMS, are shown in Table~\ref{tab:expectednumbers}. The difference in selected events between the different \Ekin values can be attributed to slightly higher reconstructed dijet masses with more pion radiation. It can be seen that the dijet event selection leads to an $S/\sqrt{B}$ of 0.21 to 3.8 depending on the quirk mass, where $S$ is the number of signal and $B$ the number of background events in the SR.

\begin{table}[htbp]
\footnotesize
    \centering
        \caption{Generated (Gen.) and expected (Exp.) number of events after event selection for dijet background and different signal hypotheses  ($\Lambda '$ = 10~\GeV) in the chosen mass window, based on 140~\ifb\ of 13~\TeV\ $pp$ data. \label{tab:expectednumbers} 
    }
   {\def\arraystretch{1.2} 
\begin{tabular}{lccccccc}
    \hline\hline
     &  \multicolumn{2}{c}{\mq = 0.75~\TeV} & \multicolumn{2}{c}{\mq = 1.0~\TeV} & \multicolumn{2}{c}{\mq = 1.5~\TeV} \\
    \Ekin &  0.2 & 1.0 & 0.2 & 1.0 & 0.2 & 1.0 \\
     \hline\hline
     Gen. Signal & 1.4e5 & 1.5e5 & 2.1e4 & 5700 & 2.3e4 & 2.3e4\\
\hline
Exp. Signal & 1.3e4 & 1.3e4 & 1800 & 1900 & 78 & 79\\
\hline\hline
Gen. Bkg & \multicolumn{2}{c}{5.4e7} & \multicolumn{2}{c}{7.9e6} & \multicolumn{2}{c}{6.5e5}\\
\hline
Exp. Bkg & \multicolumn{2}{c}{1.2e7} & \multicolumn{2}{c}{2.3e6} & \multicolumn{2}{c}{1.4e5}\\
\hline\hline
Exp. $S/\sqrt{B}$ & 3.6 & 3.8 & 1.2 & 1.3 & 0.21 & 0.21\\
\hline
\hline
    \end{tabular}
    }
\end{table}

Figure~\ref{fig:kin_distributions} shows a number of reconstruction-level distributions, comparing the simulated dijet background with the quirk signal, where \mq is chosen to be 1.5~\TeV\ and \Ekin is varied between 0.2 and 1.0. The top figure shows the invariant mass of the two leading jets. The resonant structure can clearly be seen, as well as the shift to higher invariant masses when significant pion radiation is included (\Ekin = 1.0). The remaining figures show track properties, where the tracks include both the jet tracks and the tracks from the pion radiation to avoid issues related to infrared safety: the number of tracks \Ntracks, which increases dramatically with larger \Ekin, the absolute value of the polar angle $\theta$ shifted by $\pi/2$, which encapsulates the higher centrality of the pion radiation in signal events compared to the QCD background, the transverse sphericity \cite{Sas:2021pup}, which reflects the isotropy of the radiated pions (a value of 0 corresponds to a pencil-like event, and a value of 1 to a spherical event), and  
\deltaRsquared, which describes the average distance between each track pair in the event. The higher density signal events feature tracks that are closer together than what is observed in the QCD background. Decreasing the quirk mass reduces the energy scale of the signal events and with that the track multiplicity and therefore the discrimination power of the track-based observables.  

\begin{figure}
\centering
  \includegraphics[width=0.36\textwidth]{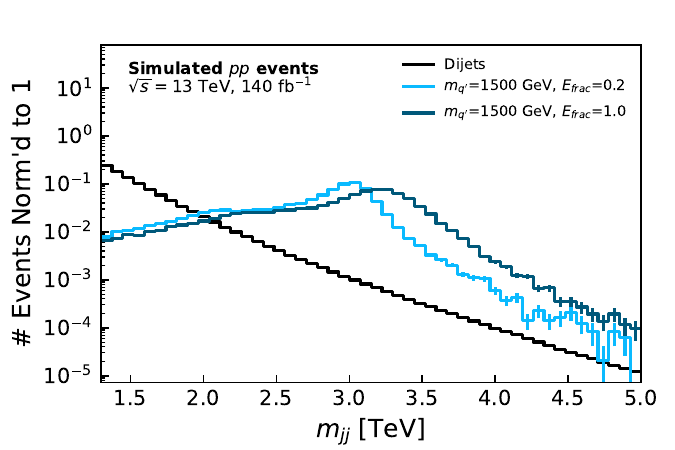}\label{subfig:invmass}
  \includegraphics[width=0.36\textwidth]{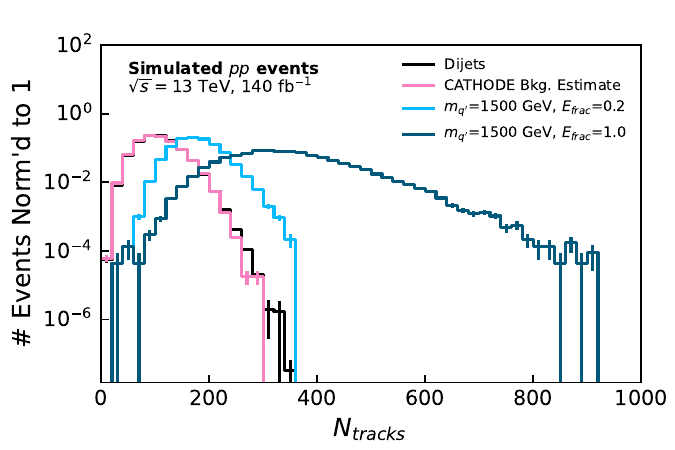}
\includegraphics[width=0.36\textwidth]{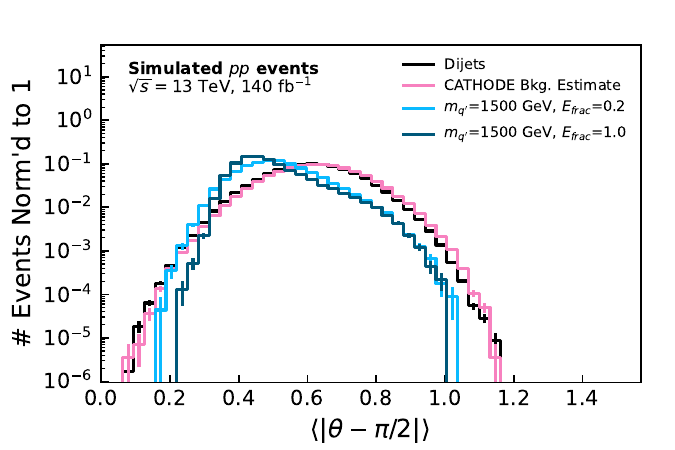}
 \includegraphics[width=0.36\textwidth]{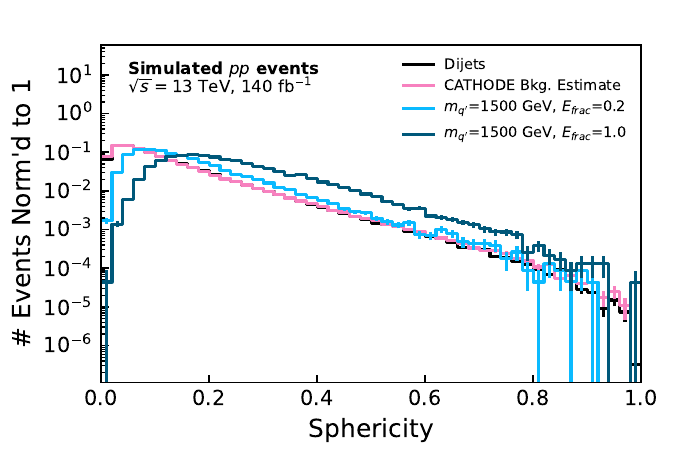}
\includegraphics[width=0.36
\textwidth]{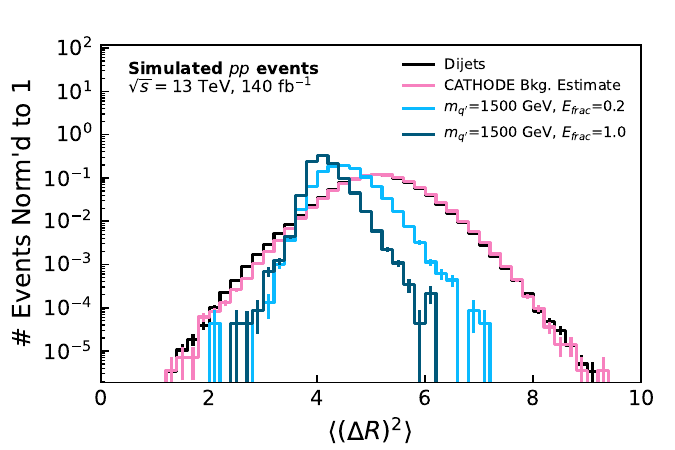}
    \caption{Reconstruction-level distributions, comparing the shape of the simulated dijet background (black) with
the quirk signal (blue), where \mq is chosen to be 1.5 \TeV\ and \Ekin is varied between 0.2 and 1.0. Also shown are the dijet background distributions estimated by a CATHODE run (pink). The invariant mass of the dijet system is shown on the top, while the other figures show track distributions in the $m_{jj}$ signal window. 
 \label{fig:kin_distributions} }
\end{figure}

\subsection{Constraints from existing searches}

\label{sec:constraints}

So far, to our knowledge, no search for resonances combined with anomalies in the low-\pt\ track distributions has been performed at any collider. However, thanks to the presence of the resonance, ``inclusive searches", i.e. bump hunts over a falling background spectrum after a dijet selection, should provide some sensitivity. 

An example is the search by the ATLAS collaboration for dijet resonances, using approximately 140~\ifb\ of 13~\TeV\ $pp$ data produced by the LHC~\cite{ATLAS:2019fgd}. 
With our simulated dijet background events and Gaussian signals, we can reproduce the expected ATLAS cross-section limits for narrow-width signals to within 40\%. The main difference comes from our imperfectly modelled dijet background. 
In Ref.~\cite{ATLAS:2019fgd}, the observed fiducial cross-section limits are given for Gaussian signal shapes with widths up to 15\%.
Comparing the corresponding excluded fiducial cross sections to the cross sections and acceptances of our model, quirk masses between 750 and 1000~\GeV\ (depending on the probed \Lambdaprime) can be excluded. This mass exclusion range is about 10\% stronger than what we present as ``inclusive'' in our results in Sec.~\ref{sec:results}, due to multiple reasons: the difference between our simulated background distribution and the ATLAS dijet spectrum, our simplified signal extraction (cut-and-count instead of a fit, see Sec.~\ref{sec:selection}), and differences between the Gaussian and our actual signal shape.

General SUEP searches (see e.g. Ref.~\cite{CMS:2024nca}) do not apply to the presented models with de-exciting QCD-charged quirks for two reasons. First, they usually target signals arising from a dark QCD sector with quasi-conformal running, which produces a SUEP of dark hadrons with some mass and effective temperature. Since these dark hadrons decay to SM final states, the associated mass scale is~$\mathcal{O}(\GeV)$ or larger. The signature discussed here arises from the direct isotropic quasi-thermal emission of SM pions, such that the resulting ``SM-QCD SUEP'' covers a different parameter space, $\mathcal{O}(100~\MeV)$, than the one covered by general SUEP searches. 
Second, even if these searches were sensitive to the discussed signal, they do not include the high mass resonance requirement, which greatly increases the sensitivity.

\section{Selection strategies}
\label{sec:selection}

This study evaluates the power of the different selection strategies by approximating the discovery sensitivity by $S/\sqrt{B}$ based on simulated signal and background events and their associated truth labels in the chosen dijet invariant mass window. Of course, in an actual data analysis, a bump hunt would be performed that involves fitting analytical functions or templates for the signal and background components to the resulting dijet invariant mass spectrum and systematic uncertainties would be included. 

\subsection{Rectangular selection cut}

The simplest way to improve the sensitivity over a resonance-only search is to apply a selection cut to the most powerful discriminant observable, which is the number of tracks (found via ablation studies using the CATHODE approach). This observable also has the advantage of being very general - many models with track anomalies will exhibit an increased number of tracks, be it from dark shower decays or oscillation radiation. 

The best choice of cut depends on the model. 
In the following, we show results for a relatively loose background efficiency of 2\%, which works better for models with less radiation, and for a background efficiency of 0.05\%, which is tighter and benefits models with many tracks in the final state, as it improves the signal$-$background separation.

\subsection{Supervised classifier}
\label{subsec:supervised}

With the primary goal to understand the maximally achievable sensitivity to the benchmark model, a supervised classifier was trained on the signal and background distributions.

The considered observables are the number of tracks, the absolute value of $\theta$ shifted by $\pi/2$, the transverse sphericity, and  \deltaRsquared, as described in Section~\ref{subsec:presel_dis} and shown in Fig.~\ref{fig:kin_distributions}.

A neural network is implemented in \textsc{pytorch}~\cite{pytorch} containing three hidden layers with 64 nodes each and a binary cross-entropy loss.  It is trained for 100 epochs with a batch size of 128,
using the Adam~\cite{kingma2014adam} optimizer with a learning rate of 10$^{-3}$. 
Half of the generated signal and background events in the signal region are used for training, 1/6 for validation, and 1/3 for testing and determination of the results. Ten trials are made, with the events shuffled between the training, validation, and testing samples. The results presented in Section~\ref{sec:results} are based on the average of these ten trials. 

 To understand the dependence on the variations within the benchmark model, the supervised classifier was also applied in a mismatched way - trained on a model with a chosen \Ekin value and evaluated for a model with another \Ekin value. 

\subsection{Anomaly detection with CATHODE}

Anomaly detection strategies are employed to be as model-agnostic as possible. They allow finding regions of phase space where a selected set of observables show differences between data (background plus possible signal) and the expected background, the latter ideally estimated purely from data control regions.  

The anomaly detection strategy chosen here is a weakly supervised search with CATHODE~\cite{Hallin:2021wme}.
CATHODE is based on weak supervision~\cite{Metodiev:2017vrx}, a training paradigm in which a classifier is trained to distinguish between a background estimate of probability density $p_\text{bkg}^\text{est}$ and data, which may contain a small fraction $\epsilon$ of signal events and has the probability density $p_\text{data} = (1-\epsilon) p_\text{bkg} + \epsilon p_\text{signal}$. If $p_\text{bkg}^\text{est}\approx p_\text{bkg}$, the classifier learns the likelihood ratio $p_\text{data}/p_\text{bkg} = 1 - \epsilon + p_\text{sig}/p_\text{bkg}$, which is monotonically related to the optimal discriminant $p_\text{sig}/p_\text{bkg}$ between signal and background.

To obtain the background estimate, CATHODE starts by splitting the sample into a SR and sidebands based on a known feature, here chosen to be the dijet invariant mass $m_{jj}$. The SRs are constructed  as described in Section~\ref{sec:simulation} using the $m_{jj}$ distributions for the 
$E_\text{frac}=1$ models.
 The conditional probability density $p_\text{bkg}^\text{est}(x|m_{jj})$ of the background in the sidebands is learned using a normalizing flow~\cite{germain2015made} conditioned on $m_{jj}$, trained using events from the sidebands.
The flow architecture and training uses \texttt{sk\_cathode}~\cite{skcathode} with the \textsc{Pyro}~\cite{bingham2018pyro} back-end and the default settings. In a second step, the flow is sampled for values of $m_{jj}$ obtained from a kernel density estimate~\cite{KDE} of the data in the signal region. These background samples are then compared to the data using a weakly supervised classifier.

The considered features are the same as in the supervised case. Different classifiers and ensembling strategies were tested. The presented results are based on a neural network with the same architecture and hyperparameters as listed above for the supervised search. 
Unlike in the supervised case, which is trained based on large simulated samples, the training in anomaly searches relies on the limited number of events available in the respective dataset. To understand the statistical spread, for every tested signal mass and every probed $\mu$, ten trials are made. The results in Section~\ref{sec:results} are the average of these trials, while the uncertainty bands correspond to the 1$\sigma$ ranges. 
Each trial involves sampling the expected number of events for 140~\ifb\ of data from the generated signal and background events.
In each trial, two non-overlapping samples are drawn. Half of the first sample is used for training, 1/6 for validation, and 1/3 for testing.
The second independent sample is used in its entirety to extract the results. 

In Section~\ref{sec:results}, results for two versions of the above are presented: Results using a so-called idealized anomaly detector (IAD), where it is assumed that the background estimate is perfect (here, the background estimate is then simply taken from a dedicated independent background sample), and the full CATHODE chain, which includes the normalizing flow estimate of the background in the signal region, as shown in Fig.~\ref{fig:kin_distributions}.

\section{Results}
\label{sec:results}

As stated above, the results in this section are based on simulated samples, assuming 140~\ifb\ of $pp$ collisions at 13~\TeV, which approximates the parameters of the LHC Run-2 dataset. Setting the stage, for the inclusive resonance search, the sensitivity $S/\sqrt{B}$ versus the signal strength is shown in Fig.~\ref{fig:sicmuinclusive} for different quirk masses and two values of \Ekin. The sensitivity depends linearly on the signal strength through the number of signal events $S$, with varying slopes due to the expected signal and background cross sections. The effect of increasing the value of \Ekin is very small and can be attributed to a small shift in the dijet peak to higher values, leading to a slightly increased $S/\sqrt{B}$ (see also Table~\ref{tab:expectednumbers}). It can be seen that  discovery sensitivity ($S/\sqrt{B} = 5$) is almost reached for the nominal benchmark model ($\mu = 1$) at a mass of 750~\GeV, while the other masses require higher signal strengths. 

\begin{figure}[h]
\centering
  \includegraphics[width=0.4\textwidth]{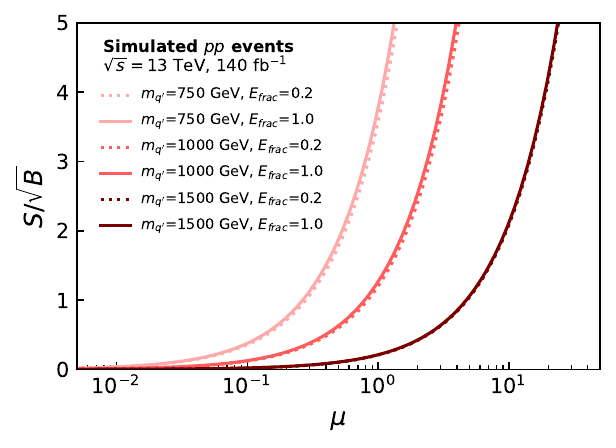}
    \caption{Sensitivity $S/\sqrt{B}$ for different signal strengths $\mu$ after the preselection, presented for \Lambdaprime = 10~\GeV, different quirk masses and two values of \Ekin. \label{fig:sicmuinclusive}}
\end{figure}

\subsection{Significance improvement curves}

A selection of significance improvement curves with respect to the inclusive resonance search is shown in Fig.~\ref{fig:sics}. The curves are calculated as a function of the false positive rate (that is, the background efficiency) for a quirk mass of 1.5~\TeV\ and two choices of \Ekin (0.2 and 1.0), as well as the different additional selection strategies described in Section~\ref{sec:selection}. The uncertainty bands are derived by sampling the background and signal samples ten times. 
The signal strengths chosen for the anomaly detection curves correspond to an \SoverB = 1 and 2 after the preselection; the other classifiers are independent of this value. 

\Ekin = 1 is a striking signature with a lot of emitted pions modifying the track multiplicity distribution in particular. All selection strategies perform very well, with a slight decrease in the simple cut-based selection compared to the multivariate methods. For both \Ekin = 1 and \Ekin = 0.2, the significance improvement in CATHODE is worse than in IAD, due to the additional challenge of estimating the background through the normalizing flow, especially in the tail of the track multiplicity. We leave the improvement of the normalizing flow in the tails of observables to future work and point out that the performance difference is less significant in the discovery potential and exclusion limits probed below. Tail-based anomaly detection has also been shown to work for the models probed in Ref.~\cite{Bickendorf:2023nej}. 

\Ekin = 0.2 is a relatively challenging signature, adding fewer additional tracks. As expected, the supervised classifier achieves the best improvement, as it is multivariate and tailored to the specific signal model. Interestingly, a mismatch of \Ekin in the training compared to the test signal does not lead to a significant performance decrease, making the supervised classifier a very attractive option for this class of models. Both anomaly detection variants lose power if the signal strength becomes too small, as the classifier is trained to discern $S+B$ from $B$, which becomes challenging for small signals. The rectangular selection cut on the track multiplicity shows a significance improvement of 2 $-$ 3, and, like the supervised classifier, has the advantage to perform independently of the signal strength, but the disadvantage of a certain model dependence.

\begin{figure}[h]
\centering
  \subfigure{ \label{subfig:sic10} \includegraphics[width=0.4\textwidth]{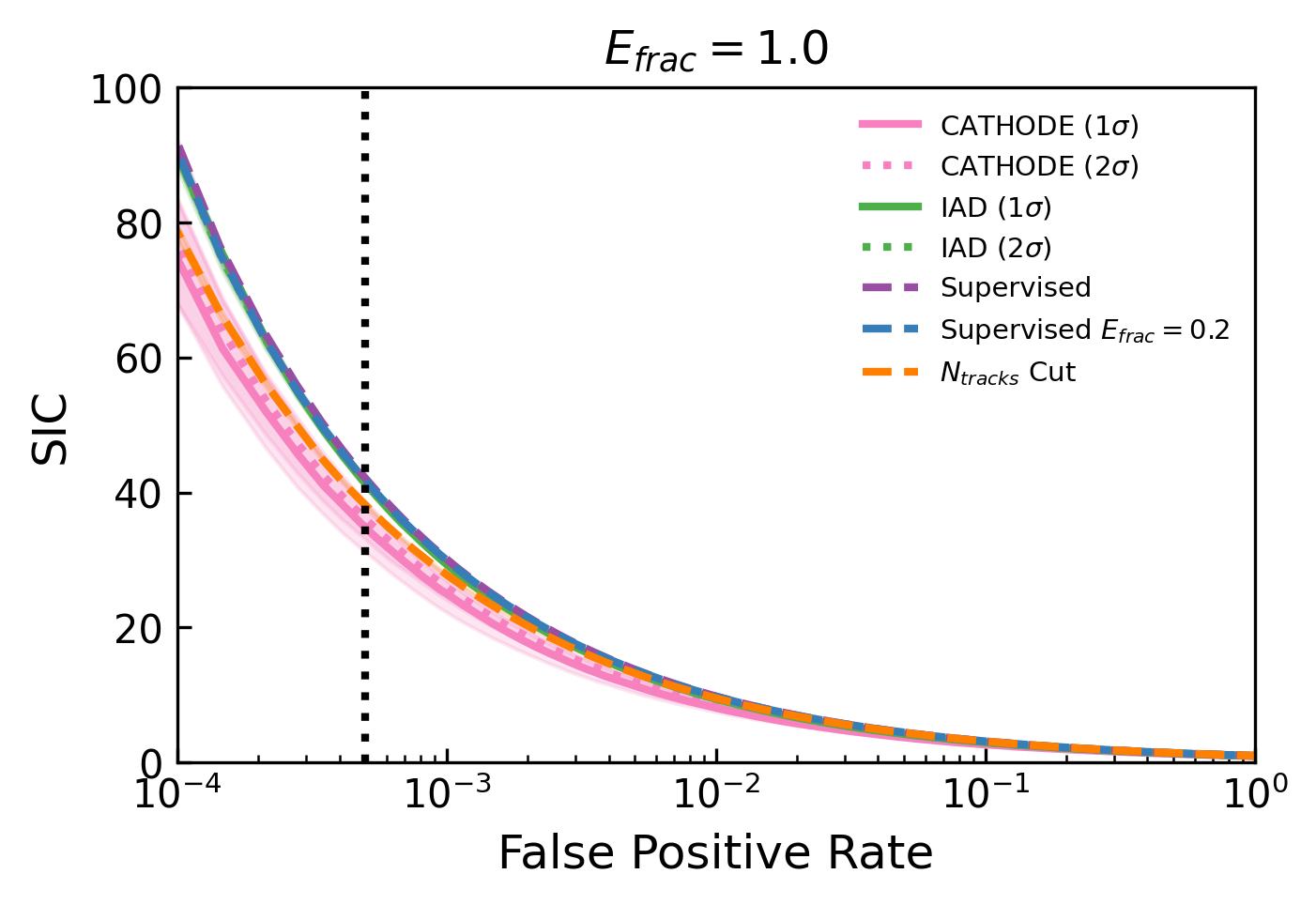} }
   \subfigure{ \label{subfig:sic02}
   \includegraphics[width=0.4\textwidth]{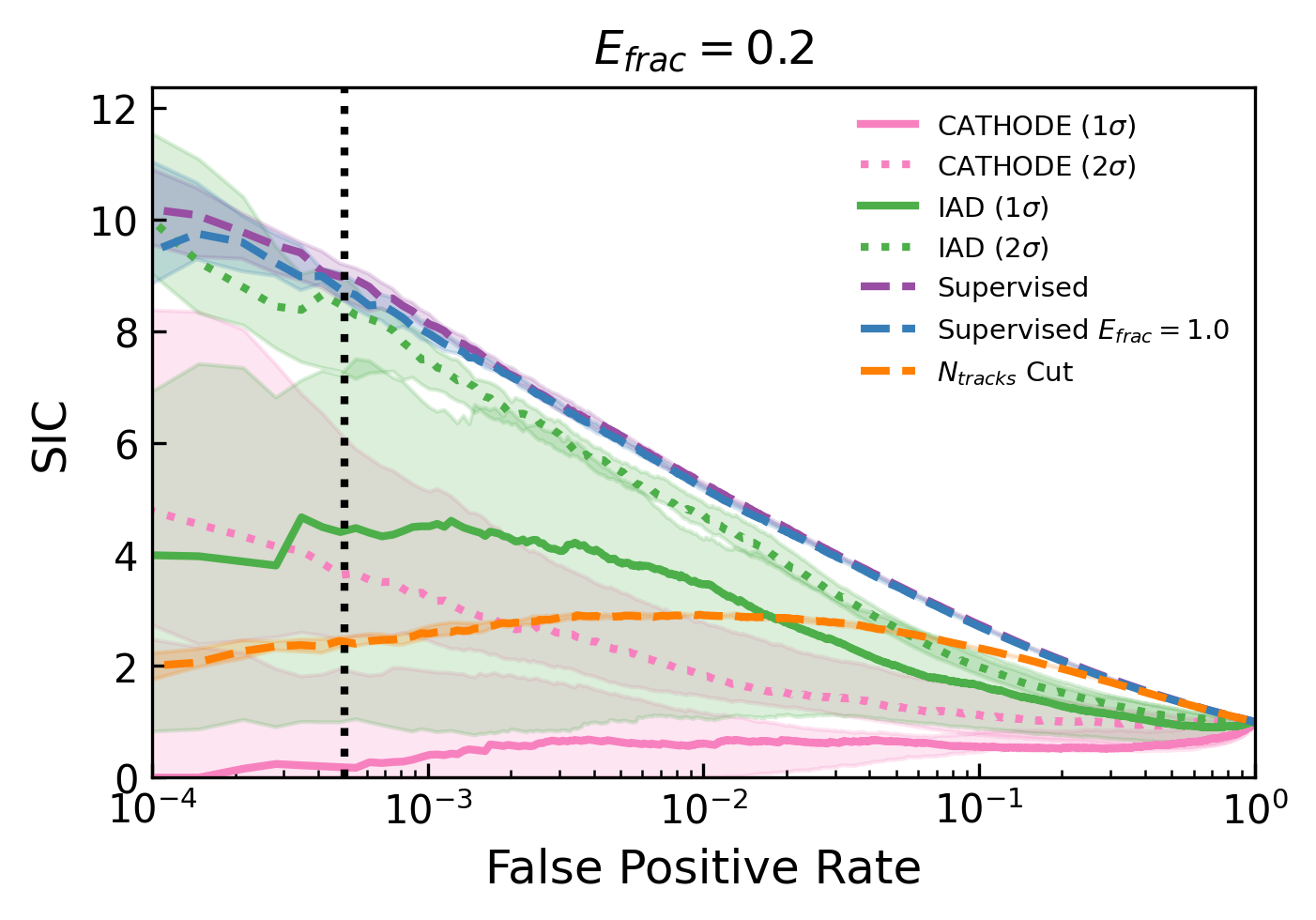} }
   \caption{Significance improvement curves for a quirk mass of 1.5~\TeV\ as a function of the false positive rate, for the different selection strategies and two values of \Ekin. The significance improvement for CATHODE and IAD depends on the signal strength; results for $S/\sqrt{B} = 1$ and 2 at preselection are presented. The uncertainty bands are derived from ten different samplings of the background and signal samples. The vertical line indicates the chosen false positive rate.  \label{fig:sics}}
\end{figure}

\subsection{Discovery potential} 

In the end, the relevant quantities are not the improvements in significance, but the detection potential, that is, the required signal strength for a 5$\sigma$ excess, and, in the absence of a signal, the exclusion limits that can be set. For these evaluations, an operating point needs to be chosen, which is usually done by picking a background efficiency. Since all selection strategies, except for the track multiplicity cut, exhibit a rising improvement with a tighter cut (see Fig.~\ref{fig:sics}), we show the results for a relatively tight background efficiency of 0.05\%. For the track multiplicity cut, we also show the results for a looser background efficiency of 2\%.

Figure~\ref{fig:discovery} shows the required signal strengths for a 5$\sigma$ discovery for different quirk masses and different values of \Ekin, employing the different selection strategies discussed in Section~\ref{sec:selection}. For the anomaly detection methods, the ranges are again the 1$\sigma$ bands from the ten samplings. 
All strategies show significant improvements over the inclusive resonance search, except in small regions of phase space around \Ekin $\approx$ 0.2. The nominal cross sections of the benchmark model ($\mu = 1$) can be probed for all tested masses, if \Ekin is high enough. 
CATHODE performs worse than the inclusive search for \mq = 750~\GeV\ and \Ekin $\approx 0.2$, as the challenging signature reduces the classifier to a random selection, decreasing the $S/\sqrt{B}$ value at a given operating point.
As expected, the supervised classifier requires the smallest signal strength for discovery. The performance decreases only minimally when the classifier is trained with a mismatched \Ekin. The rectangular cut on the track multiplicity with a background efficiency of 0.05\% performs very well, with required discovery signal strengths that are higher by $\lesssim 2$ compared to the supervised classifier. The looser track multiplicity cut improves the performance only minimally at low values of \Ekin.
For the chosen benchmark model, which should be representative of a whole class of models featuring a resonance and a higher track multiplicity compared to the background, the IAD and CATHODE performance is very similar to the track multiplicity cut. This is consistent with the SIC curves in Fig.~\ref{fig:sics}: For example, for a quirk mass of 1.5~\TeV\ and \Ekin = 0.2, the signal strength needed for a discovery with IAD or CATHODE corresponds to significances after preselection of $S/\sqrt{B} \approx$ 1 to 1.5, which can be determined by finding the $S/\sqrt{B}$ values in Fig.~\ref{fig:sicmuinclusive} that correspond to the signal strength values read off the curves in Fig.~\ref{fig:discovery}. These are values for which the anomaly detection methods exhibit similar performance as the track multiplicity cut, also in the SIC curves. 
It was verified that the anomaly detection methods indeed correctly identify the track multiplicity as the most dominant observable and rely almost exclusively on it for low amounts of signal, which do not allow the classifiers to benefit from the more subtle observables, like the angular distributions. 
Models with additional powerful and more model-dependent observables 
benefit more from the multivariateness and improved model dependence of anomaly detection methods. 
The performance difference between CATHODE and IAD for the lowest mass point is due to the less discriminant features at low quirk masses, which make the classifier more sensitive to mismodelings of the normalizing flow. 

\begin{figure}[h]
\centering
  \subfigure{\includegraphics[width=0.4\textwidth]{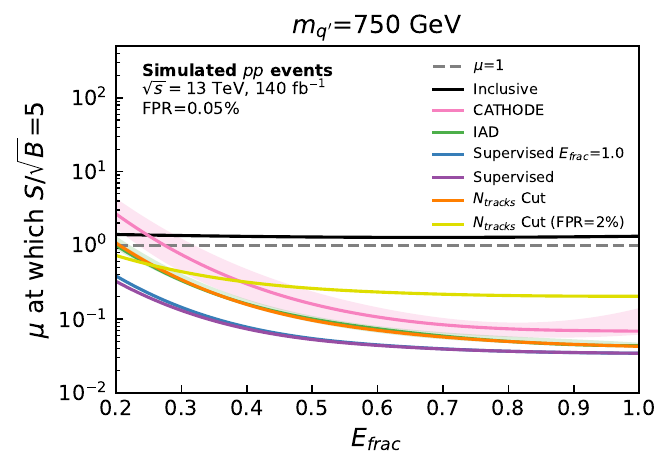}}
  \subfigure{\includegraphics[width=0.4\textwidth]{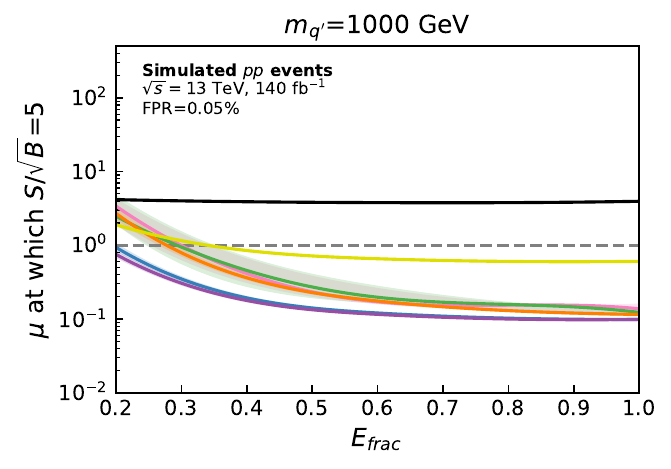}}
  \subfigure{\includegraphics[width=0.4\textwidth]{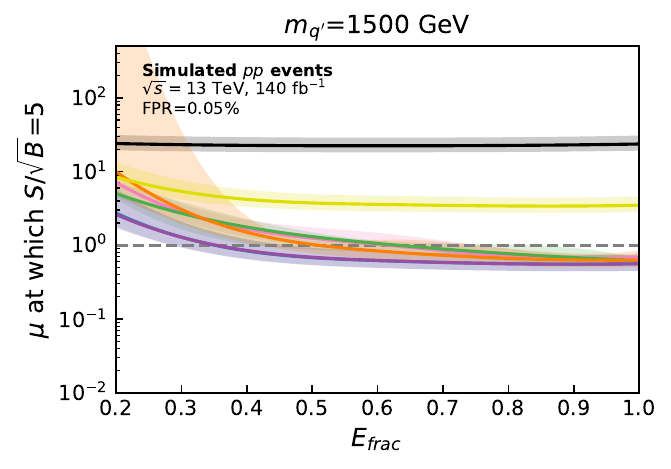}}\hspace{110pt}
    \caption{Required signal strengths for a 5$\sigma$ discovery for \Lambdaprime = 10~\GeV,  different quirk masses and different values of \Ekin, for different selection strategies (lower is better). For CATHODE and IAD, the uncertainty bands are dominated by the 1$\sigma$ spread of ten trial runs. For the other curves, they correspond to the expected statistical uncertainties on $S$ and $B$ (as the trial spread is negligible). The dashed horizontal line indicates a signal strength of 1.  \label{fig:discovery}}
\end{figure}

\subsection{Exclusion limits}

In the absence of signal, limits at the 95\% confidence limit are set as a function of the quirk mass, and different values of \Lambdaprime. The strategy followed for the limit setting with CATHODE and IAD is described in the appendix of Ref.~\cite{CMS:2024nsz}.

Fig.~\ref{fig:limits_0210} shows the exclusion limit contours for \Ekin = 0.2. It can be seen that quirk masses up to \mq $\approx$ 1450~\GeV\ can be excluded with the supervised classifier. The performance of the mismatched supervised classifier is not much worse. The loose selection cut on the track multiplicity follows, whereas IAD and CATHODE do not improve over the inclusive selection.
The worse performance of the anomaly detection compared to the discovery potential is due to the smaller values of $S/\sqrt{B}$ probed by the exclusion results. These small values make it very difficult for the classifier to find a difference between $S+B$ and $B$ even in the track multiplicity observable, as the $S+B$ distribution can be lower than $B$ due to statistical fluctuations.

As for the discovery sensitivities, \Ekin = 1 is a striking signature and applying the supervised classifier or the tight track multiplicity cut allows to exclude the whole probed phase space of \mq $< 1500~\mathrm{GeV}$ and \Lambdaprime $< 200~\mathrm{GeV}.$ The performance of IAD and CATHODE is very similar.

\begin{figure}[h]
\centering  \subfigure{\includegraphics[width=0.4\textwidth]{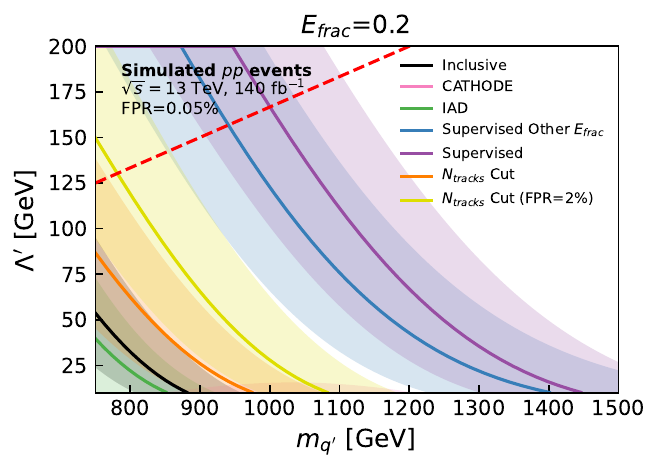}}
    \caption{Expected 95\% CL exclusion limits as a function of the quirk mass and \Lambdaprime for \Ekin = 0.2 (higher is better). The area to the left under the curves is excluded by the different search strategies.  For CATHODE and IAD, the uncertainty bands are dominated by the 1$\sigma$ spread of ten trial runs. For the other curves, they correspond to the expected statistical uncertainties on $S$ and $B$ (as the trial spread is negligible).  The  dashed red line indicates when quirk annihilation to two dark glueballs is kinematically forbidden, invalidating the naive  estimate of the branching fraction to SM jets in Fig.~\ref{fig:xs_br}. Close to the dashed red line, the quirks dominantly or entirely annihilate to SM jets, and the actual experimental limit  therefore  slightly exceeds the $m_{q'}$ mass reach found at very low $\Lambdaprime$ values. Note that the entire parameter plane is excluded  for $E_\mathrm{frac} = 1$ by the supervised classifiers and the tight track multiplicity cut.}
    
    \label{fig:limits_0210}
\end{figure}

\section{Conclusion}
In conclusion, we propose to search for a \TeV-scale resonance accompanied by anomalies in the low-transverse momentum ($\lsim$~\GeV) tracks at the LHC, thereby enhancing the discovery potential for new phenomena like de-excited bound states or dark showers. 
A benchmark quirk model and a corresponding simulation strategy were developed, representing phenomena leading to final states with a \TeV-scale dijet resonance and increased track multiplicity from radiated pions. Different classifiers relying on track information were tested. These can enhance the discovery potential of a subsequent bump hunt in the dijet invariant mass spectrum. Estimated discovery sensitivities and exclusion ranges are presented for 140~\ifb\ of 13~\TeV\ LHC data.

For the chosen benchmark model, the track multiplicity is by far the dominant discriminator, and a rectangular selection based on this observable is a simple, relatively model-independent way to enhance the sensitivity, as it does not make any assumptions on other characteristics of the new tracks. 
A supervised classifier optimized for the specific model increases the sensitivity further by making use of other, less discriminant, track properties, like angular distributions. Interestingly, the supervised classifier also performs well when an incorrect track multiplicity is assumed in the training. However, its performance could degrade if other track properties differ or are severely mismodeled. 
CATHODE, the anomaly detection method tested in this work, yields a discovery sensitivity similar to the simple track selection, due to the dominance of this observable and the small amounts of signal needed for a discovery, which prevents the method from using more granular information. The use of anomaly detection methods could, however, be beneficial to maintain a low model dependence also for more complex models with large signal-background differences in other observables. 

The suggested phenomenology covers a phase space that has not been probed at the LHC so far. It is rich and warrants further studies both by varying the type of resonance, e.g., enhancing dilepton or diphoton resonance searches, and by varying the type of particles and emission patterns of the radiation spectra. 

\FloatBarrier

\section*{Acknowledgements}
The work of D.C. was supported in part by Discovery Grants from the Natural Sciences and Engineering Research Council of Canada (NSERC), the Canada Research Chair program, the Alfred P. Sloan Foundation, the Ontario Early Researcher Award, and the University of Toronto McLean Award.
M.F.C., S.H., G.K., L.M., and M.S. acknowledge support by the Deutsche Forschungsgemeinschaft (DFG, German Research Foundation) under Germany’s Excellence Strategy – EXC2121 “Quantum Universe” – 390833306 as well as by the Bundesministerium für Bildung und Forschung under Project No. 05H24GUA. S.H. thanks the Helmholtz Association for the support through the EBP initiative. The work of D.S. was supported
by DOE Grant No. DE-SC0010008. This work has benefited from computing services provided by the German National Analysis Facility (NAF).

\bibliographystyle{ieeetr}
\bibliography{biblio}

\end{document}